\documentclass[%
 reprint,
 amsmath,amssymb,
 aps,nofootinbib,   
]{revtex4-1}
\usepackage{bm}
\PassOptionsToPackage{linktocpage}{hyperref}
\usepackage[hyperindex,breaklinks]{hyperref}
\usepackage{enumitem}
\usepackage{slashed}

\renewcommand{\theta}{\vartheta}

\newcommand{\ket}[1]{\ensuremath{\left|\, #1\right>}}

\usepackage{array}
\usepackage{mathtools}

\usepackage{etoolbox}

\begin{document} 

\title{A Microscopic Model of Holography: \\ 
Survival by the Burden of Memory}

\author{Gia Dvali} 
\affiliation{%
Arnold Sommerfeld Center, Ludwig-Maximilians-Universit\"at, Theresienstra{\ss}e 37, 80333 M\"unchen, Germany, 
}%
 \affiliation{%
Max-Planck-Institut f\"ur Physik, F\"ohringer Ring 6, 80805 M\"unchen, Germany
}%
 \affiliation{%
Center for Cosmology and Particle Physics, Department of Physics, New York University, 726 Broadway, New York, NY 10003, USA
}%

\date{\today}

\begin{abstract}  
An explicit microscopic realization of the phenomenon of holography is provided by a class of simple quantum theories of a bosonic field inhabiting a $d$-dimensional space and experiencing a momentum-dependent attractive interaction. An exact mode counting reveals 
a family of holographic states.  In each 
a set of gapless modes emerges with their number
 equal to the area of a $(d-1)$-dimensional sphere. 
 These modes store an exponentially large number of patterns within a microscopic energy gap. 
 The resulting micro-state entropy obeys the area-law reminiscent of a black hole entropy.  
We study the time-evolution of the stored patterns
and observe the following phenomenon:  Among the degenerate micro-states 
the  ones with heavier loaded memories 
survive longer 
than those that store emptier patterns.  
That is, a state gets stabilized by the burden of its own memory.
We show that due to their enhanced memory storage capacities the holographic states correspond to local minima on the pattern energy landscape. 
  When the memory load of the stored pattern is light, the landscape is flat and 
system can freely explore it.  When the load is heavy, the barriers separating the minima are steep and system gets trapped  in a local minimum for a long time until it transits into a nearest one. 
Thus, from time to time the information pattern gets off-loaded from one holographic state into another but cannot escape the system. 
During this process the pattern becomes highly entangled and scrambled.
We suggest that this phenomenon is universal in systems with enhanced memory storage capacity,  such as  black holes or critical neural networks.
This universality sheds an interesting light on the puzzle of 
why, despite the evaporation, is a black hole forced to maintain information 
internally for a very long time.  

\end{abstract}

\maketitle
\section{Holography and Lasting Memory} 
  Since the theoretical discovery of black hole entropy by Bekenstein \cite{Bek}, 
  understanding its microscopic origin has been one of the greatest challenges 
 in particle physics and gravity.  
 The area scaling of entropy gave raise to the idea of holography \cite{Hol1}. 
 It appears that the information capacity of the system is controlled by the nearly-gapless 
 quantum degrees of freedom, each inhabiting a Planck-size pixel of the horizon area. 
 The latter hypothetical modes  are often called the {\it holographic} degrees of freedom.  
  The questions that we shall focus on in the present work are: 
  \begin{itemize}
  \item  What is the microscopic mechanism behind the gaplessness of the holographic modes?  

  \item  What is the origin of the area-law scaling of their number? 
  
\end{itemize}

   For answering the above questions, it would be important to know, how general is the phenomenon of holography.  
   Usually, the holography is considered to be a property of gravitational systems. 
  Indeed, until recently, all the systems that were conjectured to exhibit this property were gravitational, 
  such as  black holes \cite{Hol1} and AdS spaces \cite{ADS1,ADS2}. 
  
   In certain cases there exist some  powerful guiding principles 
    that help in establishing the holography reliably. 
    The celebrated AdS/CFT correspondence \cite{ADS1,ADS2} is a perfect example. 
  Another important example  \cite{SV} is provided by 
  counting of black hole entropy 
  in a highly supersymmetric case with extremal black holes. Here too, the power of supersymmetry
  of string theory allow for a reliable entropy count.  
  
  However, for ordinary non-supersymmetric black holes that are deprived of  
  the above powerful tools the situation remains unclear. This is due to obvious 
  technical difficulties with non-perturbative quantum-gravity computations. Entering into these 
  difficulties is not the goal of the present paper. 
  
   Instead, following \cite{giaArea} we shall take a more modest attitude.
  The idea is to formulate a  microscopic theory of holography that 
   would be free of irrelevant quantum gravitational  technicalities. 
     At the same time the model must be powerful enough for describing the microscopic phenomenon of the emergence of the holographic gapless modes while maintaining     
  the quantum corrections fully 
     under control. 
   
 In  \cite{giaArea} it was shown that a class of simple $d$-dimensional quantum theories exhibits
 the phenomenon of holography in the above sense.   
  Namely, these theories posses a family of critical states with emergent gapless modes.  Remarkably, the number of 
latter modes  scales as the area of a sphere of one dimension less than the original space.  Consequently, the corresponding  micro-state entropy scales  as the area 
  of this  $(d-1)$-dimensional sphere.   
   These emergent gapless modes can legitimately be  
  called the {\it holographic} degrees of freedom, since they 
 fully control the sharply enhanced memory 
 storage capacity of the system.  
 Namely, the Fock space formed by these degrees of freedom 
 can store an exponentially large number 
 of patterns within a microscopically-narrow energy gap.    
   We shall refer to this Fock space as the { \it memory space}.

  In the present paper we set the following goals. 
   
  First, we would like to investigate the 
  presented class of holographic  theories and understand some aspects of their time-evolution. In particular, we wish to find out how a loaded pattern affects the time-evolution of the 
  pattern-storing state.  We shall observe an interesting phenomenon indicating that 
a state is longer-lived when the stored pattern 
uses up a larger fraction of the memory space.  Putting it shortly: \\ 

{\it  The burden of memory stabilizes the system.}  \\
 
 The  physics underlying this phenomenon can be briefly summarised as follows.    
  First, we shall explain later that the holographic states represent the local minima on the  
pattern energy landscape. This is because in each holographic state  a particular set of  
modes becomes gapless. 
Therefore,  when the system is in one of such states, a pattern can be encoded in the respective gapless 
modes at a very low energy price. 
  
 Now, when we move from a given holographic state into its nearest  neighbour,  the new set of modes becomes 
 gapless. At the same time the formerly-gapless modes gain large energy gaps. 
 Due to this,  it is energetically favourable to rewrite the pattern from the old modes into the new ones. 
 However, the road connecting the two holographic states on the energy landscape goes through a terrain where no gapless modes exist. 
This fact creates the split energy barriers that depend on the memory loads of the carried patterns.   
Namely, the barrier is higher for the patterns with the heavier memory loads. 
That is,  for carrying over a higher memory pattern, we need to climb a steeper energy slope.

  As a result, the system  is getting stuck in a local holographic state
that minimizes the energy-cost 
 of the carried pattern. 
This confinement on average lasts for a long time, until the  system can transit into a nearest local minimum.
 During this transition
 the stored pattern gets {\it off-loaded}  into a  
 set of the gapless modes  that emerge in the new minimum.  
 Thus, the original pattern gets redeposited into the modes of the new holographic state.    \\

\begin{figure}
 	\begin{center}
        \includegraphics[width=0.52\textwidth]{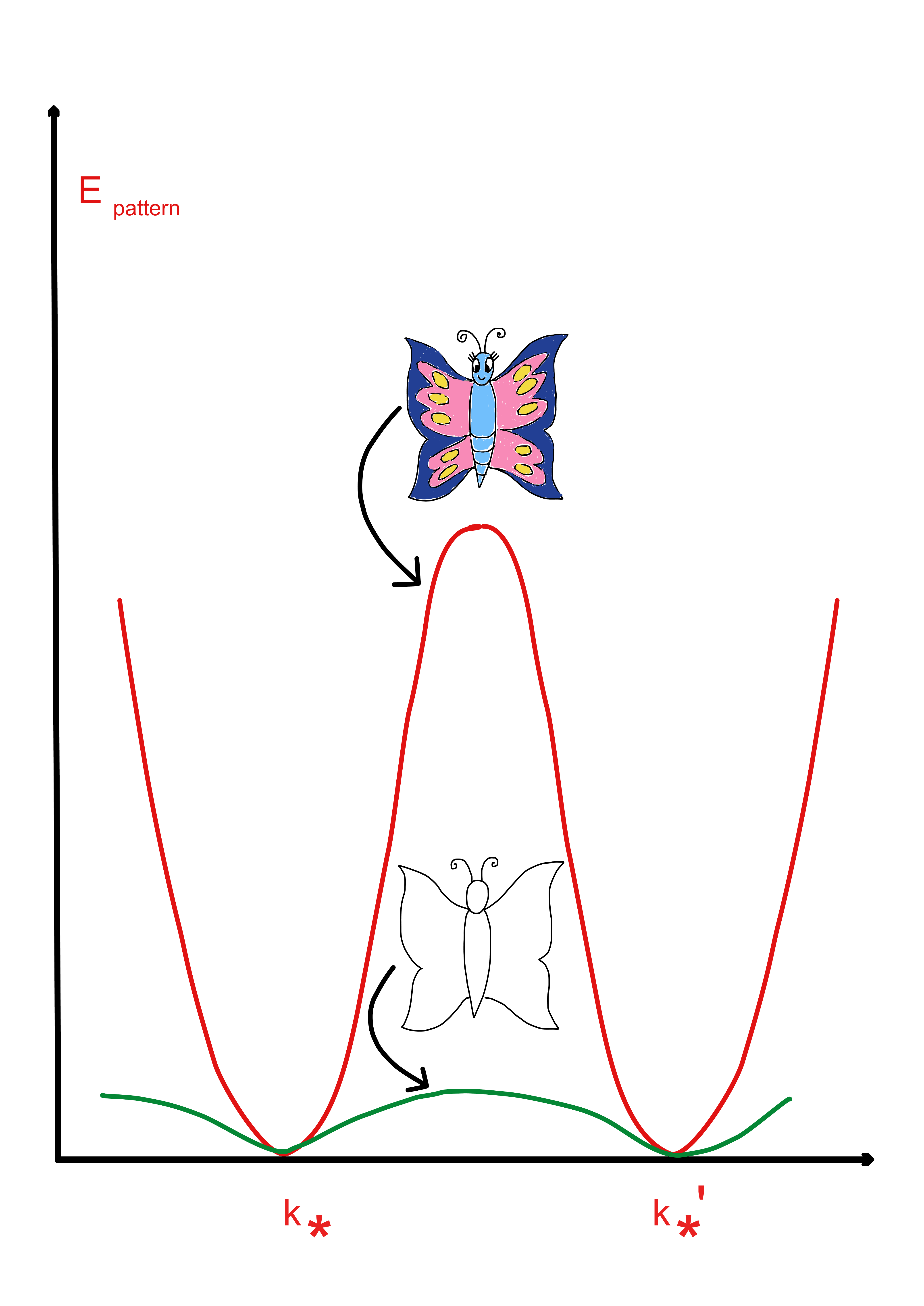}
 		\caption{A schematic description of the energy landscape 
corresponding to two patterns with high and low memory loads depicted 
 as images of a butterfly with and without the detailed features respectively.
 The corresponding energies are plotted in red and green colores.    
 The local minima, labelled by $k_*$ and $k_*'$, correspond to two distinct  holographic states, with two different sets of the gapless modes.  Although the two patterns 
 are nearly degenerate when the system is in either of the holographic states, the energy splitting 
becomes very large once we move away from such states. 
  The highly-loaded pattern (butterfly with the colored features) creates a high energy barrier and stabilizes the system in a given holographic state for a long time.
   No such obstruction occurs in the case of an emptier pattern
  described by the featureless butterfly. In the latter case the barrier 
  is shallow and the system can freely explore the landscape.
   }  
\label{landscape}
 	\end{center}
 \end{figure}

 A schematic  description of the above situation is 
 given on Fig.\ref{landscape}, where the energy landscapes  created by two patterns carrying very different memory loads are compared. 
      \\

 The time-evolution of the system  thus represents a cascade of transitions between the different holographic states with alternating sets of the emergent gapless modes.  
During this evolution the loaded pattern gets translated from one set of modes to another. In this way an  each newly-emergent generation of the gapless modes  ``inherits" 
the pattern from the formerly-gapless ones. 

  This recurrent rewriting of the pattern
    has the two major consequences. First, it is  slowing down the decay process dramatically. Secondly,  during
 its rewriting the pattern becomes scrambled, whereas the modes that store it 
 become highly entangled.

  The observed phenomenon appears to be very general and is expected to 
 take place in other systems in which the storage of information is controlled by the emergent  gapless  degrees of freedom at criticality. 
   We are not aware of analogous phenomenon previously being pointed 
   out either in the context of black holes or in other systems of enhanced memory capacity, such as the quantum neural networks. 
  Therefore, it would be natural to look for similar effects in such systems. 
 In particular, the holographic neural networks constructed  in \cite{neural, neural1} are ready made for this purpose, since 
 they share the same mechanism of enhancement of the memory storage capacity as the current model.  \\

  The second goal of the present paper is to
 search for possible connections with black hole physics. Namely, we would like to hypothesize  that a version of the presented  
  mechanism is responsible for a black hole holography.   Such a connection would shed a light at the 
 remarkable stability of the information storage by black holes from a completely  new angle.    

     Of course, we are not able to provide 
  any reliable proof from the gravity side of the story, due to the obvious technical difficulties. 
  However, we shall establishing some clear parallels between the  phenomena of increased memory storage capacity in black holes and 
   in the present microscopic model. 
   
    Regardless whether the striking similarities in the holographic behaviour of the two systems 
    are fundamental or represents a coincidence,  the existence of explicit microscopic models of holography is highly informative. 
     First, it allows to monitor the  emergence of holography at the  
  microscopic level and study its aspects within a calculable model.   
      Secondly,  it opens up a prospect of simulating such systems in the laboratory experiments with cold bosons.   
        Finally,  holographic 
     properties exhibited by the neural networks with gravity-like connections
 \cite{neural, neural1} can point towards an universal 
  underlying  mechanism of enhanced memory storage capacity in these seemingly remote systems.  \\

 \section{A Model}   
      
   We start by introducing a bosonic field operator $\hat{\psi}(\theta_a)$ defined on a $d$-dimensional sphere
   with angular  coordinates $\theta_a, ~a=1,2,...,d$, the volume element $d\Omega$ and the total volume $\Omega$. 
  We shall not specify its physical meaning, as it can be 
 applicable to different physical situations.    
For example, $\hat{\psi}$ may describe the fluctuations of
a particle number density of non-relativistic bosons on
 a $d$-dimensional sphere.  
 
   We can represent this field as an infinite sum over  the creation and annihilation operators for the modes with different generalized angular momenta, 
  \begin{equation} 
      \hat{\psi} = \sum_{k} \, Y_k(\theta_a) \hat{a}_k \,, 
     \label{expansion} 
   \end{equation}   
 where $Y_k(\theta_a)$ are the spherical harmonic functions on $S_d$. The
 label $k$ stands for a set of $d$ integers
 $k \equiv (k_1,...,k_d)$, which satisfy
 $|k_1|\leqslant k_2 \leqslant ...\leqslant k_d = 0,1,...,\infty$. 
 That is, $Y_k \equiv Y_{k_1,...,k_d}$. 
 These spherical harmonics form a complete orthonormal set. Thus,  among other properties,  they satisfy: 
 \begin{equation} 
 \int d\Omega  \, Y_k^*Y_{k'} = \delta_{kk'} \, ,
  \label{ort} 
  \end{equation} 
  where, $\delta_{kk'} \equiv \delta_{k_1k_1'}\delta_{k_2k_2'}...\delta_{k_dk_d'}$. 
  They represent the eigenfunctions of the covariant  Laplace operator, 
   \begin{equation} 
 \Delta_{\theta}   Y_k\,   = - \omega_k  Y_{k}\,, ~{\rm 
 where}~ \omega_k \equiv  k_d(k_d + d-1) \,.
  \label{eigen} 
  \end{equation} 
For our discussion it is crucial that the level $\omega_k$ exhibits the following mode-degeneracy, 
\begin{equation} \label{deg}
 {\mathcal N}_k =  \sum_{k_{d-1}=0}^{k_{d} }  \sum_{k_{d-2}=0}^{k_{d-1}}...\sum^{k_2}_{k_{1} = - k_2} \sim  (k_d)^{d-1}\,.
   \end{equation}

 The operators $\hat{a}_k^{\dagger}, \hat{a}_k$, are the 
 creation and annihilation operators for the eigenmodes
 corresponding to the eigen-number-sets $k$. They satisfy the usual    
 algebra, 
       \begin{equation}   \label{algebra} 
    [\hat{a}_j,\hat{a}_k^{\dagger}] = \delta_{jk}\,, \, \, 
  [\hat{a}_j,\hat{a}_k]  =   [\hat{a}_j^{\dagger},\hat{a}_k^{\dagger}] =0\,.  
 \end{equation}
 
We are now ready to introduce a simple microscopic model of holography. 
We shall construct a model that represents a small modification of the prototype model presented in \cite{giaArea} 
and is described by the following Hamiltonian, 
 \begin{eqnarray}   \label{BH} 
  &&  \hat{H}  =  \int d\Omega \Big\{ 
  -  \hat{\psi}^{\dagger} \Delta \hat{\psi} \, - \,  
    {2\Omega \over \Lambda} (\hat{\psi}^{\dagger} \Delta \hat{\psi}^{\dagger}) 
   ( \hat{\psi} \Delta \hat{\psi}) 
    -  \\ \nonumber
   && - \, {\Omega^2 \over 2 \Lambda^2}  \, \left ( (\hat{\psi}^{\dagger 2}  \Delta^2  \hat{\psi}^{\dagger})
    (\hat{\psi}^2 \Delta  \hat{\psi}) \, + \, (\hat{\psi}^{\dagger 2}  \Delta  \hat{\psi}^{\dagger})(\hat{\psi}^2  \Delta^2 \hat{\psi} )
 \right ) \Big\} \,.    
 \end{eqnarray}
Here $ \Delta \equiv e \Delta_{\theta}$
and $e$  is an unit of energy. 
For example, if $\hat{\psi}$ were to describe a 
fluctuation of a number density of a non-relativistic particle 
of mass $m$, we could take $ e = {\hbar^2 \over 2mR^2}$.  \\
 
 Let us consider the states in which only the zero momentum mode 
 is macroscopically occupied, i.e., the states $\ket{}$ that satisfy $\hat{a}_0^{\dagger} \hat{a}_0 \ket{} = N \ket{}$,
 with $N$ large.   At the same time the occupation numbers of other modes are assumed to be small.  How small, shall be quantified shortly.

  As  it is well-known, on such states, using  Bogoliubov approximation
  \cite{bogoliubov}, we can replace the  operators by  $c$-numbers:  $\hat{a}_0^{\dagger} =  {\rm e}^{-i\alpha} \sqrt{N}, ~
  \hat{a}_0  =  {\rm e}^{i\alpha} \sqrt{N}$, where $\alpha$ is an unimportant  phase.    
 
 Next, in order to make the mode-counting exact, we take the  following double-scaling limit
   \begin{eqnarray}  \label{doublescale1}
 &&N \rightarrow \infty\,,
~~  \Lambda \rightarrow \infty, \\ \nonumber
 && {e \over \Lambda} N^{3d \over 3d -1} = {\rm finite} \,.
 \end{eqnarray}
 The meaning of the very last relation will become clear in a moment.   
 In the above limit all the terms that are high order in creation/annihilation operators 
vanish and the Hamiltonian takes a simple form,
   \begin{eqnarray}  \label{HAMILTON} 
 &&\hat{H}  =   \sum_{k\neq 0}^{\infty}  \epsilon_k  \hat{a}_k^{\dagger}\hat{a}_k, \\ \nonumber 
&& {\rm where,} \\ \nonumber 
&&  \epsilon_k=  e\omega_k \, \left(1 - {N e\over \Lambda} \omega_k\right )^2  \, .
 \end{eqnarray}
 Of course, in reality we can take $N$ to be arbitrarily large and finite.
 From (\ref{HAMILTON}) we observe that the levels that satisfy
 \begin{equation} \label{NKA}
 N k_d(k_d+d-1) = {\Lambda \over e}  \,, 
 \end{equation}
  become {\it gapless}. 
  Notice, from (\ref{NKA}) and (\ref{doublescale1}) it is clear that 
  corresponding $k_d$ scales as, 
   \begin{equation} \label{Kscaling}
 k_d \sim N^{{1 \over 6d -2}}   \, . 
 \end{equation}
   The expression (\ref{Kscaling}) is telling us 
 that with increasing $N$ the gapless level (\ref{NKA})  is pushed towards higher angular harmonics. 
  
 Now, the number of the emergent gapless modes is equal to the corresponding level-degeneracy given by (\ref{deg}).  For large $k_d$ this scales as 
 \begin{equation}  \label{area} 
   {\mathcal N}_k  = \left ( {R \over L_k} \right )^{d-1},   
 \end{equation} 
 where  $L_k \equiv \hbar \sqrt{{N \over 2\Lambda m}} = 
 \hbar \sqrt{{1 \over 2em\omega_k}}$.  We thus observe that the number of the 
 emergent gapless modes,  ${\mathcal N}_k$,  scales as the area of a $(d-1)$-dimensional sphere in units
 of the scale $L$, exactly as in the model of \cite{giaArea}.  \\
  
  It is easy to see that the memory storage capacity of the system is almost fully controlled by the 
  modes of the critical level satisfying (\ref{NKA}), since other levels maintain very large energy gaps.  
For example,  the energy gaps of the modes of a neighbouring level $k_d' = k_d + l$, 
for $k_d \gg l$ are given by, 
\begin{equation} \label{gapA1}
\epsilon_{k'} 
\simeq  4e|l|^2 \, .
\end{equation} 
Therefore, storing information is such modes is very costly in energy.

In contrast, an information stored in the gapless modes of  the  critical level $k_d$ has a negligible energy cost. 
For example, the patterns can be stored in the Fock space spanned over the basis vectors of the form, 
\begin{equation} \label{Basic}
  \ket{N-n}_{a_0}\otimes \ket{pattern}_{k} \, ,  
  \end{equation} 
where $\ket{N-n}_{a_0}$ is a number eigenstate of the zero-momentum  mode with the eigenvalue $N-n$, 
whereas 
 \begin{equation} \label{PatternB}
 \ket{pattern}_{k}  \equiv  \ket{n_{1}^{(k)} ,....,n^{(k)}_{{\mathcal N}_k}}  
  \end{equation} 
 denote the number eigenstates of the gapless modes belonging to the critical level 
$k_d$. 
Here $n_{j}^{(k)}$ are their respective eigenvalues and the index $j =1,2,..., {\mathcal N}_k$ numbers all possible sets $k_1,k_2,...k_d$. 
  We shall refer to the Fock space spanned over the basis pattern vectors (\ref{PatternB}) as the {\it memory space}.   For definiteness, we shall restrict ourselves to the binary values $n_{j}^{(k)} = 0$ or $1$.
  This  leaves us with a $2^{{\mathcal N}_k}$-dimensional memory space.  Generalization of our analysis 
  to a larger dimensional memory  space is trivial\footnote{However, as shown in \cite{class}, 
  the states with $n_{j}^{(k)} \gg  1$ should {\it not } be included into the micro-state entropy count, 
  since they do not belong to the given micro-ensemble.}. \\
  
   It is clear that the corresponding micro-state entropy
  $S = {\mathcal N}_k {\rm ln}2$  satisfies the same area-law (\ref{area}) as 
  ${\mathcal N}_k $.  This is highly reminiscent 
  of Bekenstein-Hawking entropy of a black hole with $L$ playing the role of the Planck length.  \\

  For the future convenience we shall introduce a measure 
of the usage of the memory space in form of a ratio
$0\leqslant  {n^{(k)} \over {\mathcal N}_k} \leqslant 1$, where 
$n^{(k)}$ is the total occupation number of the gapless modes, 
 \begin{equation} \label{nG} 
 n^{(k)} \equiv \sum_{ j= 1}^{{\mathcal N}_k}\, 
  n_{j}^{(k)} \, ,
\end{equation} 
in a given pattern (\ref{PatternB}).  
Since the Hamiltonian (\ref{BH}) conserves the total particle number, 
the nearly-degenerate states of the form (\ref{Basic}) belonging  
to the same super-selection sector must satisfy 
$n^{(k)} = n$.

  The energy cost of a pattern (\ref{PatternB}) is easy to estimate by taking into account 
  the $N$-dependence of different parameters given by (\ref{doublescale1}), 
 (\ref{NKA}) and (\ref{Kscaling}). 
     Then, the contribution to the energy of the pattern (\ref{PatternB}) coming from 
  (\ref{HAMILTON}) is  
   \begin{equation} \label{gapP}
E_{Pattern}  \simeq  e \omega_{k} {n^3 \over N^2}\,.
\end{equation} 
The patterns that use the maximal capacity of the memory space are 
the ones with $n \sim  {\mathcal N}_k$. Then, taking into account that 
(\ref{deg}) and (\ref{Kscaling}) give ${\mathcal N}_k \sim N^{d-1 \over 6d -2}$,   we get that (\ref{gapP}) is maximized 
by  
   \begin{equation} \label{gapPmax}
E_{Pattern}  \sim \, e N^{-{3 \over 2}}\, .   
\end{equation} 
This is a negligible energy cost.

  However, we still need to check the contribution coming from the 
  higher order terms in creation and annihilation operators of the gapless modes. 
  So far, such terms were ignored in the Hamiltonian  (\ref{HAMILTON}) because they are suppressed 
  by powers of $N$.  Hence,  due to relatively-small occupation numbers 
  of the gapless modes, they were assumed to be negligible. 
  The consistency of this assumption must be checked by estimating the expectation value of the Hamiltonian  over the states of the form (\ref{PatternB}) in which we take  $n \sim  {\mathcal N}_k$.  
  Taking into account the scalings (\ref{doublescale1}),
 (\ref{NKA}) and (\ref{Kscaling}),   
it is easy to see that this expectation value is within $e$. 
Thus, even with the contributions of all the interaction terms included, 
a maximal energy gap occupied by $2^{{\mathcal N}_k}$ possible patterns of the form  (\ref{PatternB}) fits within the elementary gap $e$. \\ 

 To summarize,  the  energy gap occupied by the entire  $2^{{\mathcal N}_k}$-dimensional 
 memory space is narrower than the unit gap $e$.  We thus see that 
 the model (\ref{BH}) exhibits a manifestly holographic behaviour. 
 It delivers a memory space of exponentially enhanced pattern storage capacity. This space 
 is created by the emergent gapless 
 modes of the number ${\mathcal N}_k$ that satisfies the are-law (\ref{area}).

 \section{Holographic Jumps}

 We shall now investigate how a loaded pattern affects the time-evolution 
 of the ``host" holographic state.  
We focus on a process of a {\it decay} which refers to a decrease of the initial occupation number $N$
of the constituent zero-momentum particles $\hat{a}_0$. 
 Such a process may result from their conversion into the modes of a new bosonic field that plays the role of an external reservoir. 
We denote the latter by $\hat{\chi} = 
\sum_k Y_k(\theta_a) \hat{b}_k$, where  $\hat{b}_k^{\dagger},\hat{b}_k$ are its creation and annihilation 
operators that satisfy the usual algebra of the type (\ref{algebra}).  

  The mode-conversion then can be easily achieved by introduction 
of the following terms in the Hamiltonian, 
\begin{equation}  \label{EVAP}  
   \hat{H}_{\chi}  =  \int d\Omega 
\Big\{ 
 - \hat{\chi}^{\dagger} \Delta \hat{\chi} 
   + \,  {e \over 2 N^{{3d -2 \over 3d -1}}} \left (\hat{\psi}^{\dagger}  \hat{\chi} \, +  \, 
   \hat{\chi}^{\dagger}  \hat{\psi} \right )
  \Big\} \,.  
   \end{equation}   
  The first term in (\ref{EVAP}) stands for the standard kinetic energy of the 
 $\hat{b}$-particles, whereas the second one describes the mixing between $\hat{a}_k$ and 
  $\hat{b}_k$-modes.  The latter enables an exchange of the occupation number 
  among the two sectors while conserving the total number of particles.  
For an observer monitoring the $\hat{\psi}$-field the conversion 
$a \rightarrow b$ looks as a {\it leakage} of $a_0$-quanta.  
 The strength of the mixing is chosen by the requirement that it 
 should not offset 
 the gaplessness of the $\hat{a}_k$-modes
 that satisfy the condition (\ref{NKA}).
 
  Note, in (\ref{EVAP}) we have chosen a simplest interaction between the
 external modes  and the 
   holographic system that allows a decay of the latter. 
    The reader can generalize our analysis for other types of 
  decay processes.  Such a generalization shall not change our results qualitatively, as long as it respects the near-gaplessness of the holographic modes.     
  \\

Let us first find the variation of $N$ that is required for changing
the critical level from $k_d$ to its neighbour  $k_d'= k_d+ l$. 
 From (\ref{NKA}) it is clear that such a variation is
\begin{equation} \label{ChangeN}
\Delta N_l \simeq - N{2l \over k_d} \simeq - lN^{1-{1 \over 6d - 2}} \, .
\end{equation} 
Below we shall study the nearest neighbour transition $|l|=1$.

 For describing such transitions in the double-scaling regime (\ref{doublescale1}),  the relevant part of the Hamiltonian is the one containing the zero modes $\hat{a}_0$ and 
 $\hat{b}_0$ and the $a_k$-modes of the two neighbouring levels  $k_d$  and  
 $k_d +1$.
 For definiteness, we shall denote them by $k_*\equiv k_d$ and $k_*' \equiv k_d +1$ respectively.
 All the other modes possess very high energy gaps and can be ignored in our analysis.

Expressing the relevant parameters through $N$,  we can write the effective Hamiltonian in a simplified form, 
   \begin{eqnarray}  \label{H222} 
 &&\hat{H}_{eff}  =   e  N^{{1 \over 3d -1}}   \Big\{  \, \left({\hat{a}_0^{\dagger} \hat{a}_0 \over  N} -1 \right )^2 
 \sum_{k \in k_*}  \hat{a}_k^{\dagger}\hat{a}_k \, 
   +  \\ \nonumber 
 && +   \, \left({\hat{a}_0^{\dagger}
 \hat{a}_0 \over  N - \Delta N_1} -1 \right )^2 \sum_{k \in k_*'}   \hat{a}_k^{\dagger}\hat{a}_k \, 
   +  \\ \nonumber 
&& + \, {1 \over 2 N} \, (\hat{a}_0^{\dagger} \hat{b}_0 + \hat{b}_0^{\dagger}\hat{a}_0) \,  + \, ... \,
 \Big\}\,,
 \end{eqnarray}
where $\Delta N_1$ is given by (\ref{ChangeN}) for 
$l=1$ and  ``$...$" stand for the interaction terms among $k_*$- and 
$k_*'$-modes.  The latter terms shall be ignored for the time being.  As we shall see, 
for enabling the time-evolution of certain patterns these terms are important. \\

 Now, using the above effective Hamiltonian, we wish to time-evolve 
 from an initial state in which the occupation number $N$
 of $a_0$-mode satisfies (\ref{NKA}) for the level-$k_*$. The resulting gapless modes,
 $k \in k_*$, are assumed to form a pattern (\ref{PatternB}). The initial occupation numbers 
 of all the other modes are set to zero.  Thus, the initial state can be written as:
   \begin{eqnarray} \label{INstate}
 && \ket{N}_{a_0} \otimes \ket{0}_{b_0}\otimes\ket{pattern}_{k_*}\otimes\ket{0}_{k_*'} \,. 
   \end{eqnarray} 
   We are now approaching the crucial point:  The time-evolution 
 of the state vector is extremely sensitive to
 $n^{(k_*)}$.  
  
  Namely, for $n^{(k_*)} =0$ 
  the time evolution represents an oscillation between the $a_0$ and $b_0$-modes with a periodic 
 exchange of the occupation number among 
  them. For example, the expectation value of the occupation number of the  $a_0$-mode evolves 
 in time as $\langle \hat{a}_0^{\dagger}\hat{a}_0\rangle =  
 N{\rm cos}^2\left ( {t e \over 2\hbar}  N^{-{3d -2 \over 3d -1}} \right)$ and vanishes periodically. 
 
  However, for  $n^{(k_*)} \neq 0$ the situation changes dramatically and the oscillation
 becomes suppressed due to the following reason. 
   For $\langle \hat{a}_0^{\dagger}\hat{a}_0\rangle = N$
     the level-$k_*$ is gapless and the patterns 
  (\ref{PatternB}) cost very little energy. 
 However,  for $\langle \hat{a}_0^{\dagger}\hat{a}_0\rangle \neq N$ the modes of the level-$k_*$
 acquire large gaps and the same pattern becomes very costly in energy.   
 Indeed, as it is clear from (\ref{HAMILTON}) and (\ref{NKA}),  a variation 
  $\langle \hat{a}_0^{\dagger}\hat{a}_0\rangle = N +\delta N$ generates the energy gaps for the $k_*$-modes equal to 
  \begin{equation}\label{dgap} 
 \epsilon_{k_*} =  e  {(\delta N)^2 \over N^{2 - {1 \over 3d -1}}}\,. 
 \end{equation}
 Consequently, the energy cost of the pattern (\ref{PatternB}) becomes 
 \begin{equation}\label{Pcost}
 E_{pattern}  =  \langle \hat{H}_{eff} \rangle = 
   e  {(\delta N)^2 \over N^{2 - {1 \over 3d -1}}}  n^{(k_*)}\,, 
 \end{equation} 
 where the expectation value of the Hamiltonian is taken over 
 the state 
\begin{equation} \label{Wrongstate}
 \ket{N -\delta N}_{a_0} \otimes \ket{\delta N}_{b_0}\otimes\ket{pattern}_{k_*}\otimes\ket{0}_{k_*'}\,.
 \end{equation}  
 For  $\delta N = \Delta N_1$ and $n^{(k_*)}\sim {\mathcal N}_{k_*}$  this energy cost is 
 $E_{pattern} \sim e N^{{d-1 \over 6d -2}}$. This energy is macroscopically larger than the elementary energy gap $e$.  
 This high energy-cost results into a large level-splitting between the {\it would-be}  initial 
 (\ref{INstate}) 
 and final  (\ref{Wrongstate}) states 
  and kills oscillations. \\

   That is, the system is ``reluctant" to abandon a holographic state when the   
  later carries a high memory load. 
  The phenomenon that we are observing can be described as: 
  {\it  A stabilization of the system by its memory-load. } \\

   It is useful to capture the essence of the story by analysing the following  slightly simplified version of the Hamiltonian 
  \begin{eqnarray}  \label{H3} 
 &&\hat{H}_{eff}  =   e  N^{{1 \over 3d -1}}   \Big\{  \, \left(1 - {\hat{a}_0^{\dagger} \hat{a}_0 \over  N} \right )
 \sum_{k \in k_*}  \hat{a}_k^{\dagger}\hat{a}_k \, 
   +  \\ \nonumber 
&& + \, {1 \over 2 N} \, (\hat{a}_0^{\dagger} \hat{b}_0 + \hat{b}_0^{\dagger}\hat{a}_0) \,  + \, ... \,
 \Big\}\,,
 \end{eqnarray}
 for which the evolution of the state (\ref{INstate}) can be derived exactly.  
 Here the  level $k_*'$ has been ignored, since their occupation numbers are set to zero.
   
  The exact time-evolution of the state  (\ref{INstate}) gives,  
     \begin{eqnarray} \label{evolve} 
 && \langle \hat{a}_0^{\dagger}\hat{a}_0\rangle =  N \left ( 1 -  A\, {\rm sin}^2\left ( {t \over \tau} \right) \right ) \, ,
 \end{eqnarray} 
where 
  $A \equiv { 1 \over  1 + (n^{(k_*)})^2}$
 and  
 $\tau \equiv  {2 \hbar  N^{{3d - 2 \over 3d -1}} \over e \sqrt{1 + (n^{(k_*)})^2}}$.
 
  The above expression clearly shows the tendency that the large values of $n^{(k_*)}$ block the change of the occupation number 
$\langle \hat{a}_0^{\dagger}\hat{a}_0\rangle$. 
 That is, the states with heavier loaded memories get stabilized, whereas 
 the lighter-loaded ones oscillate freely.

 For example, let us compare the two extreme cases.  For the state $n^{(k_*)}=0$ that corresponds to an empty  pattern $\ket{pattern}_{k_*} = \ket{0,0,...,0}_{k_*}$, 
 we have $A =1$.
 Thus, the number of $a_0$-particles fully diminishes after 
  the time $t={\pi \over 2} \tau$. 
   
  In contrast, for  $n^{(k_*)} = {\mathcal N}_{k_*}$ that represents a fully loaded pattern $\ket{pattern}_{k_*} = \ket{1,1,...,1}_{k_*}$, we have $ A \simeq   N^{-{d-1 \over 3d -1}}$.  
 This means that for $d> 1$ the number of $a_0$-particles essentially cannot change. \\

The essence of the story is clear.  Each holographic macro-state that satisfies (\ref{NKA}) represents a state of exponentially enhanced memory storage capacity. The gapless modes that emerge around this state can store large patterns in their memory at a negligible energy cost. 
  Although different patters use vastly different fractions of the memory space, the energy difference among them is negligible as long as 
  we are at the given critical (holographic) point.  Once we start changing $N$ and move away from the holographic  state, the energy splitting between the patterns increases 
  according to (\ref{Pcost}).  
      In this respect, the variation 
  $\delta N$ plays the role analogous to an external magnetic field that produces a spectral line splitting  
 in Zeeman effect \footnote{We thanks Cesar Gomez for suggesting this analogy}.   
  
  The patterns that use bigger fraction of the memory space  start costing more energies. 
   Due to this reason, the states with maximally loaded memory 
get ``frozen" until they manage to ``off-load" the pattern into a next 
 holographic level.  Thus,  our initial state can evolve towards the nearest  holographic 
 state as long as the pattern is
 off-loaded  from the level-$k_*$ into the level-$k_*'$. 
  That is, the transition has the following form, \\

    \begin{eqnarray}  \label{transit} 
 \ket{N}_{a_0} \otimes \ket{0}_{b_0} &&\otimes\ket{pattern}_{k_*}\otimes\ket{0}_{k_*'} \\ \nonumber
 && \downarrow \\ 
 \ket{N -\Delta N}_{a_0} \otimes && \ket{\Delta N}_{b_0}\otimes\ket{0}_{k_*}\otimes\ket{pattern'}_{k_*'}\,. \nonumber
  \end{eqnarray}  \\
  
    A cartoon description of the energy landscape is given by Fig.\ref{landscape}. Here, the patterns encoding images of colorless and colorful butterflies
 use up smaller and larger values of $n^{(k_*)}$  respectively.  Correspondingly,  the colorful butterfly creates a higher energy barrier separating the two holographic states of (\ref{transit}). \\

  The above implies the following two things. \\   
  
  First, rewriting the pattern from one 
 set of modes to another requires an involvement of the 
  interaction terms among $k_*$ and $k_*'$-modes. These are 
  suppressed by powers of $N^{-1}$. This suppression slows down the transition 
  process  and prolongs  the life-time in a given 
  holographic state. \\
   
\begin{figure}
 	\begin{center}
        \includegraphics[width=0.53\textwidth]{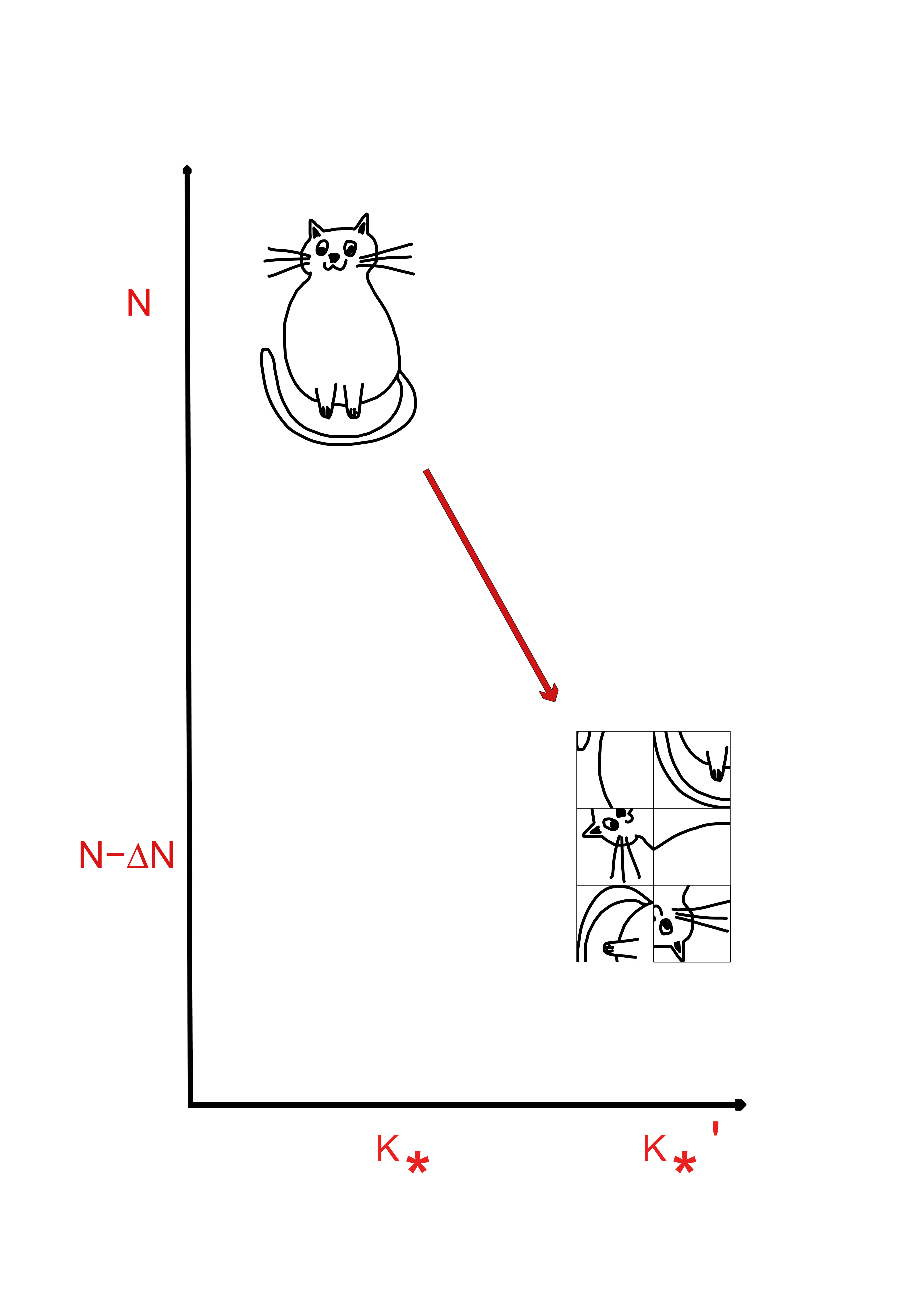}
 		\caption{A schematic description of the transition (\ref{transit}) in $N-k_*$ plane. The original pattern is represented by an image of a cat and the final one by its scrambled version.  When extending the analogy to a black hole the transition process can be viewed as the change of the black hole state due to the Hawking evaporation in which  $\hat{b}$-quanta are impersonating an outgoing Hawking radiation. During this process black hole state moves from one holographic level to another while the stored pattern is off-loaded into the new modes and becomes scrambled.  The fraction of information that is leaking into the $\hat{b}$-sector 
is suppressed by powers of ${1 \over N}$ and by energy splitting 
among the modes and is negligible. 
Thus, during its decay process
 the holographic system maintains the stored pattern for a long time while reloading it in a scrambled form. Although the analogy only serves the purpose of qualitative understanding of the process, 
it extends surprisingly far. } 
\label{transition}
 	\end{center}
 \end{figure}

   Secondly, in the process of rewriting the 
 pattern necessarily becomes {\it scrambled} in the following sense. 
  Assume that a pattern is stored originally in a basis ket-vector of the form
 (\ref{PatternB}) that represents a number eigenstate of the gapless modes of the level $k_*$.  
 During the transition (\ref{transit}) the pattern is erased from the level-$k_*$ and rewritten in the level-$k_*'$.  This process involves minimum $n$ destruction operators 
 of the level-$k_*$ and the same number of creation operators 
 of the level-$k_*'$, i.e., is generated by an effective operator involving 
 $n$ 
 $\hat{\psi}^{\dagger}$-s and the same number of  $\hat{\psi}$-s. 
However,  due to the rotational invariance of the Hamiltonian (\ref{BH}), a given set of destruction operators of the level-$k_*$ is coupled to a complicated superposition of creation operators 
 of the level-$k_*'$. As a result, the pattern $\ket{pattern}_{k_*}$ becomes off-loaded 
 into a superposition of many different number-eigenstates of the level-$k_*'$. 
 In other words, the final pattern $\ket{pattern'}_{k_*'}$  in (\ref{transit}) - into which 
 the original pattern $\ket{pattern}_{k_*}$ is rewritten - will not consist of a single basis pattern 
 of the form (\ref{PatternB}).  Rather, it will contain a superposition of large number of basis patterns,
 \begin{equation} \label{scr}
  \ket{pattern'}_{k_*'} = \sum_{n_j} c_{n_1n_2...}\ket{n_1,...,n_{{\mathcal N}_{k_*'}}}\, ,
  \end{equation}
  with different coefficients $c_{n_1n_2...}$. 
 Obviously, in such a state the modes of the newly-emerged gapless level $k_*'$ are expected to be highly entangled.  
 A schematic description of such a transition process  is
presented in Fig.\ref{transition}. 
       \\

 We finish this section by stressing that the observed phenomenon should not be confused with an effect of stabilization of 
 a given macro-state due 
 its high micro-state entropy \cite{class}. 
 Instead, we are observing a  dramatic difference in time-evolutions 
 of the micro-states belonging to the same macro-state. For example, 
 the patterns $\ket{0,0,....,0}_{k_*}$ and  $\ket{1,1,....,1}_{k_*}$ belong to the ensemble describing one and the same macro-state. They are degenerate in energy and contribute equally to the micro-state entropy count. 
 Nevertheless, their time evolutions are extraordinarily different, due to the fact that they use very different fractions of the memory space. \\
 
 Let us ask,  how probable are the patterns with high memory-loads? From simple combinatorics we conclude that at large-$N$ they are maximally probable, since the number of patterns is peaked around 
  $n_{k_*}\sim {\mathcal N}_{k_*}$. Namely, for large ${\mathcal N}_{k_*}$
  the number of patterns is maximized for  $n_{k_*}= {1 \over 2} {\mathcal N}_{k_*}$ and is equal to $\begin{pmatrix}
      {\mathcal N}_{k_*}   \\
        {1 \over 2} {\mathcal N}_{k_*} 
\end{pmatrix} \sim {2^{{\mathcal N}_{k_*}}\over \sqrt{{\mathcal N}_{k_*}}}$. Thus, 
in such patterns 
we can encode a maximal amount of information. 
In the same time, the relative numbers of patterns 
with smaller values of $n_{k_*}$ are exponentially suppressed and 
vanish for ${\mathcal N}_{k_*}\rightarrow \infty$. 
 Thus, if the choice of an initial pattern is random, with maximal probability 
 we are going to encounter a state from the highest memory bunch.  As we have seen these are very long lived.

\section{Black holes}

 We now wish to establish some parallels between the information storages in a black hole 
 and in the above-discussed non-relativistic holographic model. 
  For this we shall compare the thought experiments
  performed by the two observers, Alice and Bob.  
  Both observers are instructed to maximize the energy-efficiency 
 of the pattern storage in a box of a radius 
  $R$.  
   The difference is that Alice works with gravity, whereas Bob works with a non-relativistic quantum field model (\ref{BH}).

 Consider Alice first.  The number of patterns that Alice can store in the box can be maximized by 
 storing them in the excitations of the modes with minimal wavelengths.  
 Indeed,  the number of localized modes with a certain 
 microscopic wave-length $L$ 
 scales as the volume of the box,  ${\mathcal N} \sim \left({R \over L}\right)^{d}$.  We shall impose a cutoff given 
 by the Planck length, $L=L_P$. 
 Thus, a naive estimate of the number of
  modes that can store patterns is ${\mathcal N} \sim\left({R \over L_P}\right)^{d}$. 
 
 However,  it is not solely the number of modes that matters but also the
 energy required for their activation.    In fact, it appears that Alice cannot
 actualize even a tiny fraction of modes without 
 conflicting with gravity.
  Indeed, in the absence of interactions, the energy threshold of excitation for each Planck-wavelength mode would be  $\sim {\hbar \over L_P}$.  Consequently, the energy of a pattern that would exploit an order-one fraction of such modes would be, 
   \begin{equation}\label{EP1}
  E_{pattern}  
    \sim  {\hbar \over R} \left({R \over L_P}\right)^{d+1}\, .    
  \end{equation} 
   However, it is simply impossible to concentrate this amount of energy in a box of size $R$, without running in conflict 
 with black hole physics.  
 Indeed,  using a well-known relation between the   
mass of a black hole and its radius $R$, 
\begin{equation}\label{Mass} 
 M_{BH}  \sim {1 \over R} \left({R \over L_P}\right)^{d-1}\, , 
  \end{equation}  
we can rewrite (\ref{EP1}) as 
 \begin{equation}\label{EP}
  E_{pattern}  \sim M_{BH} \left({R \over L_P}\right)^{2}\, .    
  \end{equation} 
 Thus, the energy of a pattern 
 would exceed the mass of a black hole of radius $R$
  by a factor  $\left({R \over L_P}\right)^{2}$!

Obviously, this is absurd.  
Thus, not all the Planck momentum modes contribute into the entropy count, since they cannot be actualized simultaneously. 
The above may be taken as good news, since it seems 
to show why the entropy of a system cannot scale as its volume in Planck units.
   
 However, the above is not a fully reliable argument, since it ignores the interactions among the modes. Notice, using the 
same naive reasoning, Alice would encounter 
 an analogous problem also with reproducing the area-law entropy. 
 Indeed, the energy of non-interacting Planck-wavelength particles with the occupation number
 equal to the area of the box still hugely exceed the mass of an $R$-size black hole, 
   albeit  only by a factor  $\sim \left({R \over L_P}\right)$. Again, this is 
absurd.

 The only reliable conclusion of the above discussion is that 
 a naive mode-count that ignores  their gravitational influence on each other does not give a proper understanding of 
 the information storage capacity of the system.  
 Something highly non-trivial must happen 
 to the modes in order to accommodate a black hole equivalent 
 of the information capacity while not exceeding the black hole energy.

  Precisely this is what happens when a black hole is formed. So,   
 Alice can use a trade-off.   
  Namely, Alice first needs to convert  the box into a black hole
by investing a moderate energy given by (\ref{Mass}).  After this is achieved, she can take advantage of  
the resulting micro-state degeneracy for storing patterns 
within a negligible energy gap. 
   
   In order to form a black hole,  Alice can populate the box by 
  soft gravitons.  These have the wavelengths $\sim R$, and correspondingly, the energies of order $1/R$. 
 It is easy to estimate \cite{NP} that after placing roughly $N \sim \left({R \over L_P}\right)^{d-1}$
  of such soft particles in the  box the relation between the mass  of the 
  box  ($M_{Box} \sim N/R$) and its size ($R$) becomes that of a black  hole 
 (\ref{Mass}). This signals that the box became a black hole. Now,  due to its usual Bekenstein-Hawking 
 entropy, $S \sim \left({R \over L_P}\right)^{d-1}$,  the black hole houses an  
exponentially large number of nearly-degenerate patterns in form of its micro-states. 
  The  holographic interpretation that Alice would give to the observed enhancement of the memory storage capacity of the system is in terms of emergence of some mysterious gapless modes.

  Thus, from Alice's perspective a ``miracle" has 
  happened: After populating a soft mode to a certain critical occupation number
 $N$, a set of ${\mathcal N}$ gapless species emerged.  Plus, the entropy tells her 
that  the number of modes  
 ${\mathcal N}$  scales as the area. \\
 
  Alice has no precise knowledge of the microscopic mechanism behind the appearance of the gapless species. 
 However, one thing is clear:  
 They emerge in a critical state in which a soft mode is macroscopically occupied. 
  Notice, in this emergence there is a striking similarity with the non-relativistic system considered by Bob.
  This similarity may tell us - at least qualitatively - what is the origin of Alice's gapless modes.  \\

   Therefore,  let us follow Bob's experiment.  He too is trying to store a maximal number of patterns within a minimal energy gap. However, unlike Alice, he is  using a bosonic field 
 $\hat{\psi}$ on a $d$-dimensional sphere of radius $R$. 
  For designing his Hamiltonian Bob borrows some minimal guidelines 
  from Alice.  Namely, he endows bosons with {\it gravity-like} momentum-dependent attractive interaction that is described by the second term in the Hamiltonian (\ref{BH}).   It is intriguing that with this extremely naive way 
 of introducing ``gravity" in the system, the story repeats itself 
 in one to one correspondence with Alice's case.    \\

     Indeed, here too, in the absence of an attractive interaction among the bosons, a storage of patterns in the 
   box
  would be very costly 
   in energy. For example, a typical pattern of the sort 
 (\ref{PatternB})
   would cost energy,  
  \begin{equation}  \label{BEP}
   E_{pattern} \sim  e\omega_*  k_*^{d-1} = e  \left ( {R \over L_{k_*}} \right )^{d+1}\, .
  \end{equation}  
  This is macroscopically larger than the elementary gap $e$. 
  Putting aside the obvious particularities of a non-relativistic 
 system,  the problem is qualitatively very similar to the one encountered by Alice.  This is already clear from comparing (\ref{BEP}) and (\ref{EP1}).

  However,  just like Alice in the previous example, Bob can  
  bypass this difficulty 
 by first populating the softest mode 
  to a critical occupation number $N$ given by (\ref{NKA}).  At this point the excitation energy gap of $k_*$-modes collapses to zero and the patterns can be stored extremely cheaply.   
  However, unlike Alice, Bob has an advantage of working with a microscopic theory in which
 the number and the origin of the gapless mode species can be explicitly traced. Thanks to the  
  double-scaling (\ref{doublescale1}) the quantum corrections are under control and the mode-counting is made exact.   
 This makes the origin of the gapless modes fully transparent. As we have already discussed in details, 
 they emerge because of negative energy interaction.  
 This interaction makes sure that an 
 increase of the occupation number of the soft mode $a_0$  lowers the excitation thresholds for the 
 ``hard" modes.
 This leads to the family of critical occupation numbers (\ref{NKA}) for which the corresponding sets of 
 the $k_*$-level modes become gapless.   The number of the gapless species obeys the area-law 
 (\ref{area}) and they represent the holographic degrees of freedom.
\\

  With the above knowledge, we are now ready to hypothesize that the emergence of the holographic gapless 
  degrees of freedom in Alice's case is of a similar nature.  Namely, 
 by forming a black hole Alice effectively 
 creates a multi-particle state with high occupation number of 
 soft gravitons \cite{NP}. 
 The key point is that due to the gravitational coupling among the modes,  an increase of the  occupation number of the ``soft" gravitons
  lowers the excitation energy thresholds of the ``hard" ones. The role 
  of these hard modes in  Alice's  case is presumably played by the Planck wavelength quanta.     
  For a certain critical occupation number $N$ of the soft modes the energy gap of the hard ones is expected to collapse to zero.  It is natural to assume that this is when the black hole state is reached.  \\
  
 Our interpretation thus would be that a black hole represents a state of high occupation number $N$ of soft gravitons \cite{NP} at a quantum critical point  \cite{DG}, exactly as advocated by the ``black hole $N$-portrait". \\
 
 Next, using the following dictionary 
  \begin{itemize}
  \item  soft master mode $\hat{a}_0$  $\rightarrow$  soft constituent gravitons 
  \item  gapless modes  $\hat{a}_{k_*}$    $\rightarrow$  hard gapless gravitons 
  \item  $\hat{b}_0$ -mode  $\rightarrow$ soft particles of Hawking radiation
  \item  $\hat{b}_{k_*}$-modes $\rightarrow$  hard free particles  
\end{itemize}
Alice and Bob can qualitatively explain in the language of the presented toy holographic model some peculiarities of black hole radiation. 
Namely, they can answer the question: \\

{\it  Why, despite emitting a radiation, is the black hole forced to maintain a stored information for a very long time?} \\

Here is how the model explains this. 
The black hole stores an information pattern of the type (\ref{PatternB})  in the hard gravitons (the analogs of $\hat{a}_{k_*}$-modes). These are gapless due to a critical occupation number $N$ of the soft ones (the analogs of $\hat{a}_0$).   
Now, the black hole decays due to a conversion of these  soft gravitons 
(the analogs of $\hat{a}_0$) into the quanta of Hawking radiation of the similar softness (the analogs 
 of  $\hat{b}_{0}$).  
  Due to this process $N$ decreases by $\delta N$ and the pattern becomes costly in energy (the analog of (\ref{Pcost})),
 since the level $k_*$ acquires a small but non-zero gap $\epsilon_{k_*}$
(the analog of  (\ref{dgap})).  This gap however is not nearly as large 
 as to match the 
 energies of the free quanta of the same momenta $k_*$ (the analogs of $\hat{b}_{k_*}$).
  Due to this, there is no chance to release 
 the pattern into the outgoing quanta of Hawking  radiation.  Indeed, such a process  
 would require a conversion  $\hat{a}_{k_*} \rightarrow \hat{b}_{k_*}$ which 
 is extremely suppressed due to the difference in energy gaps.  
  Recall that  due to criticality  the energies  of the holographic modes $\hat{a}_{k_*}$ are negligible
  as compared to the energies of the free particles $\hat{b}_{k_*}$  of the same momentum $k_*$.  
  That is, there are no external modes available for off-loading the pattern carried by the black hole.  
  This is the reason why the black hole is forced to maintain 
 the pattern internally and translate it into the modes of a newly-emerged holographic level 
 $k_*'$. During this translation the pattern becomes highly scrambled.  \\
 
  In addition, the whole process is slowed down by the burden of memory $n_{k_*}$ of the carried  pattern. 
  This fact allows to make a curious prediction.  
 Namely,  the  two black holes that in the classical description appear identical
 in quantum theory may acquire different life-times. These lifetimes would be determined by their memory loads $n_{k_*}$.  
 
 Of course,  the relative number of patterns with low memory loads
 ${n_{k_*} \over {\mathcal N}_{k_*}} \ll 1$  is exponentially 
 suppressed and vanishes in the limit of infinite $N$. 
  This explains why the above phenomenon has not been 
 noticed previously in semi-classical analysis of black holes. Indeed,  the semi-classical limit 
 corresponds to $N = \infty$. Therefore,   
the probability to encounter a semi-classical black hole carrying a low-memory pattern is zero. 
  Thus, if our analogy holds, the mechanism of stabilization by the burden of memory should be operative practically for all large black holes. 
However, the effect may become statistically significant for the small ones. \\

  In establishing the parallels between the two systems, it is important 
  to distinguish the unessential differences from the more fundamental ones. 
 For instance, an unessential difference is the choice of a soft ``master" mode.  The soft modes that Alice is using for reaching the critical state have the finite momenta $\sim {1 \over  R}$, whereas Bob can get away 
 by using a zero momentum mode.  This is an artefact of periodic boundary conditions of Bob's system which allow him to populate a finite size box with a zero momentum mode.
 This difference can be easily removed by replacing a $d$-sphere with 
 a $d$-dimensional ball.   It is evident that the analogs of the presented gaplessness mechanism persists both for  
 $d=1$ \cite{1Dnonperiod} as well as for $d>1$ cases \cite{3Dnonperiod}.  
 
 A more fundamental difference 
 is of course the freedom that Bob has in the choice of the Hamiltonian. This is expected to be much more restrictive for gravity. \\
  
  Of course, it would be too naive to expect  that a simple non-relativistic system can capture 
  all the aspects of the black hole information storage.   
  However, the remoteness  of the two
   systems makes their similarities even more unexpected. 
   Whether these similarities are manifestations of a real underlying connection remains to be understood.  
  Independently of this question,  the existence of the class of theories with explicit microscopic origin 
  of holography is useful from several perspectives. \\
    
  In particular, such models provide a simple theoretical laboratory
  for addressing the number of fundamental questions and discovering the new phenomena. 
  In the present paper we have identified such an effect by 
  observing that the states with heavier loaded memories decay slower.    
 This phenomenon is sufficiently general and is expected to  extend beyond the considered class of toy models. \\
 
   Our results suggest that the long term internal maintenance of information is a generic property of the holographic systems. It originates from an inability to off-load a pattern stored in the gapless modes into 
any other external quanta due to a very high energy difference.   \\

  As a separate question for future studies, it would be natural to investigate what is the connection, if any,  between the phenomenon that we 
 described as the {\it scrambling of the pattern}  in the process of its off-loading and the
  black hole fast scrambling conjecture \cite{fast}. 
  An explicit microscopic mechanism of scrambling was identified in \cite{scrambling}.
 There it was observed that the
 fast scrambling and chaos are exhibited already by a simplest one-dimensional model of attractive bosons 
 on a ring. The crucial ingredients were:  A high-density of states and a Lyapunov exponent
 describing the classical instability of transition towards this set of states.  This model
 shares a  similarity with the presented one in the sense of existence of a critical state with a gapless excitation.  However, there are crucial differences.  First, in \cite{scrambling} the number of gapless modes is too small \footnote{However, see an example in the second reference in \cite{scrambling} where also the higher momentum modes are made gapless
 at criticality.  The scrambling  has not been studied for such a case.}  and the area-law holography cannot be defined. Secondly, in the present case the system exhibits a family of 
holographic states  and can experience ``jumps" among them.   
  Thus, the logical step would be to apply analysis of \cite{scrambling} to the presented class of the holographic models. 
 The detailed numerical studies of the scrambling process during the holographic jumps of the type 
 (\ref{transit}) will be given in \cite{numbers}.   \\

 As concluding remarks,  the phenomenon of stability of a state by its memory load  
   generalizes to other systems in which the enhanced memory capacity
   is due to gaplessness, such as the neural networks  introduced in 
   \cite{neural,neural1}.  Finally,  it should be possible to simulate the models discussed in the present paper in table-top experiment with cold bosons \cite{atoms}.  \\

\section*{Acknowledgements}
We thank Lukas Eisemann, Cesar Gomez, Marco Michel, Sebastian Zell for discussions and 
Tamara Mikeladze-Dvali for discussion on interdisciplinary applications.  
This work was supported in part by the Humboldt Foundation under Humboldt Professorship Award and ERC Advanced Grant 339169 "Selfcompletion".

\end{document}